# Self-diffusion in isotropic and liquid crystalline phases of fd virus colloidal rods: a combined single particle tracking and differential dynamic microscopy study†

Eric Grelet, *[a] Vincent A. Martinez [b] and Jochen Arlt *[b]

In this article, we investigate the dynamics of self-organised suspensions formed by rod-like fd virus colloids. Two methods have been employed for analysing fluorescence microscopy movies: single particle tracking (SPT) in direct space and differential dynamic microscopy (DDM) in reciprocal space. We perform a quantitative analysis on this anisotropic system with complex diffusion across different self-assembled states, ranging from dilute and semi-dilute liquids to nematic and smectic organisations. By leveraging the complementary strengths of SPT and DDM, we provide new insights in the dynamics of viral colloidal rods, such as long time diffusion coefficients in the smectic phase. We further discuss the advantages and limitations of both methods for studying the intricate dynamics of anisotropic colloidal systems.

## 1 Introduction

Brownian colloids exhibit a rich variety of dynamic behaviours, which can usually be tuned by modulating particle interactions and concentration. Among these particles, rod-like colloids, characterised by shape anisotropy, display intricate phase behaviour, giving rise to the formation of liquid crystalline organisations. This results in a more complex dynamics, which, beyond its phase dependence related to the dimensionality of the translational and orientational orders, is strongly anisotropic, having therefore distinct diffusion along and perpendicular to the rod long axis.[1] Various methods have been devised to assess the diffusion of rod-shaped particles, including dynamic light scattering,[2,3] fluorescence correlation spectroscopy,[4,5] X-ray photon correlation spectroscopy,[6] fluorescence recovery after photo-bleaching,[7,8] and single particle tracking (SPT).[9,10] However, many of these approaches have been limited to the exploration of the isotropic liquid phase, as working with liquid crystalline organisation involves monitoring sample orientation and addressing anisotropic optical properties like birefringence, leading to scarce studies in the dense self-assembled states. One promising technique is differential dynamic microscopy (DDM),[11] which has been recently successfully used to characterise dilute[12–14] as well as dense suspensions[15] of anisotropic diffusers.

Here, we aim at using and comparing differential dynamics microscopy and single particle tracking to investigate the dynamics of a model system of colloidal rod-like particles in a wide range of concentrations, from disordered liquid to lamellar liquid crystalline phases. As rod-shaped colloidal rods, we have employed the filamentous fd viruses widely used as model system in soft matter.[16,17] Such micrometer long viral rods are monodisperse in shape and size and have been shown to behave nearly as hard rods.[18–21] This biological system is therefore suitable for studying the dynamics in the different self-assembled liquid crystalline states by fluorescence microscopy using single particle tracking. Specifically, the anisotropic diffusion coefficients have been determined both in the nematic[10] and the smectic mesophases, where hopping type diffusion has been evidenced.[22–24] This approach to investigate dynamical phenomena at the individual particle scale has been found to be very efficient and fruitful; however, it also has some intrinsic drawbacks such as limited statistics as well as some limitation for probing the long times (loss of tracked particles and photobleaching of the dyes). By contrast, DDM analyses the temporal intensity fluctuations over a range of length scales to extract the dynamics of the system by providing the intermediate scattering function (ISF).[11] As there is no need to resolve individual particles it is often possible to use large fields of view to achieve high statistics. The measured ISFs may allow to identify the nature of the physical mechanisms responsible for the particles dynamics of a sample. However, quantification of these phenomena requires an explicit model of the ISFs to extract relevant kinetic or dynamical parameters. Often simple generic models are sufficient to account for complex systems such as colloidal gels,[25] but more sophisticated models enable for instance to extract detailed population averaged dynamical parameters, *e.g.* for suspensions of active swimmers.[26–28] Although it is in principle straightforward to generalise the analysis to direction dependent dynamics, DDM studies focusing on anisotropic fluids such as liquid crystals are quite scarce to date.[12,15,29,30] In this paper, both DDM and SPT methods are performed and compared by probing the whole self-organised phases of fd suspensions, from dilute and dense isotropic liquid phases to nematic and smectic states, and new insights in the dynamics of this rod system are presented.

[a] *Univ. Bordeaux, CNRS, Centre de Recherche Paul-Pascal, UMR 5031, 115 Avenue du Dr Schweitzer, F-33600 Pessac, France. E-mail: eric.grelet@crpp.cnrs.fr*
[b] *School of Physics and Astronomy, James Clerk Maxwell Building, Peter Guthrie Tait Road, The University of Edinburgh, Edinburgh EH9 3FD, UK. E-mail: j.arlt@ed.ac.uk*

† Electronic supplementary information (ESI) available. See DOI: https://doi.org/10.1039/d4sm01221f

## 2 Methods

As system of colloidal rods, we use the filamentous M13KE bacteriophage, a mutant of fd, which has a contour length $L = 1$ μm, a diameter $d = 7$ nm, and is semi-flexible with a persistence length of $L_p = 2.8$ μm,[19] resulting in a rod end-to-end distance $L_e = 0.94$ μm according to Kratky & Porod model for isolated polymeric chains.[31] For some samples, we also use a stiffer mutant, Y21M, which exhibits a persistence length $L_p = 9.9$ μm and has $L_e \simeq L = 0.92$ μm.[19,32] Both viral mutants have therefore similar end-to-end length and identical diameter. They are grown using the ER2738 strain as *E. coli* host bacteria and purified following standard biological protocols. A small fraction of viruses (in the range of $1:10^5$ to $1:10^3$ depending on the experiment, see details below) is labelled with green (Alexa Fluor 488-TFP ester, Invitrogen) or red (Dylight550-NHS ester, ThermoFischer) dyes to enable their visualisation by fluorescence microscopy. Samples were imaged using a 100× NA 1.4 UPLSAPO objective on an IX71 Olympus microscope equipped with a NEO sCMOS camera (Andor) having a pixel size of 13 μm in binning 2, and a Omicron LedHub as light source. Time series (33 fps) are recorded for different dilutions of virus suspensions initially dialysed against a Tris–HCl–NaCl buffer at pH 8.2 and ionic strength $I = 20$ mM, ranging from isotropic liquid, to (chiral) nematic and smectic mesophases. In the liquid crystalline phases, the samples exhibit a planar orientation between cover slip and glass slide, as illustrated in Fig. 1. The same set of optical fluorescence data is used for analysis by single particle tracking (using in-house designed Matlab codes) or by differential dynamics microscopy (using codes developed with LabView).

In the next paragraphs, we will give an outline of the two methods compared in this manuscript, as they are both well established in literature.[33,34]

### 2.1 Single particle tracking (SPT

The experimental conditions are optimised for single particle tracking analysis: only a small fraction ($1:10^4$ to $10^5$) of tracers, *i.e.* viruses labelled with fluorescent dyes (see above), are added to the sample. This allows for a good detection of uniquely distinguishable trajectories $\mathbf{r}(t)$ obtained using a custom-written particle tracking algorithm developed in MATLAB (MathWorks).[22,24,35] The mean squared displacement (MSD) at lag time $\tau$ defined as $\text{MSD}(\tau) = \langle [\mathbf{r}(\tau + t) - \mathbf{r}(t)]^2 \rangle_t$ where $\langle \ldots \rangle_t$ denotes the average over start times $t$, is calculated for each trace, before being first averaged over the total number of detected particles (a few hundreds per virus concentration) and then fitted to determine the corresponding anisotropic diffusion coefficients $D_{\parallel/\perp}$ and exponents $\gamma_{\parallel/\perp}$ according to $\text{MSD}_{\parallel/\perp} = 2 D_{\parallel/\perp} t^{\gamma_{\parallel/\perp}}$ in the direction parallel $\parallel$ and perpendicular $\perp$ to the mean particle orientation.[23] Such an approach allows for a direct and instantaneous access and visualisation of the particle anisotropy in the liquid crystalline phases.

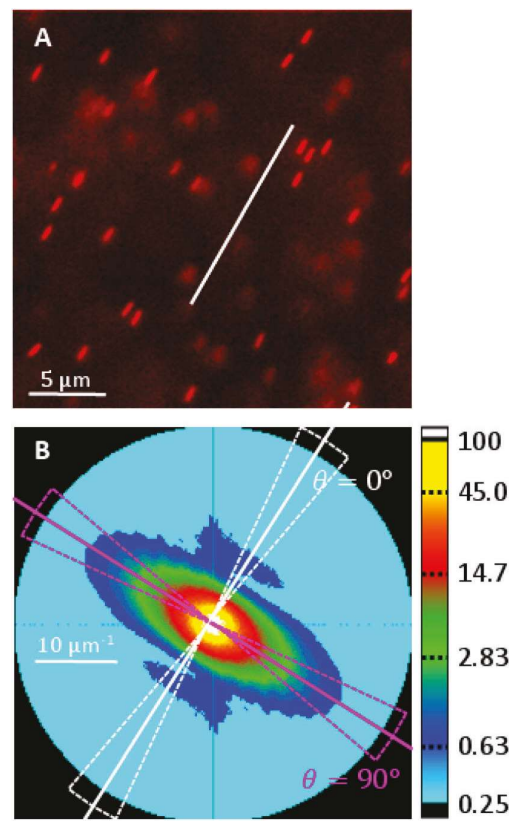

Fig. 1 (A) Snapshot showing a 26.0 μm square sub-region extracted from a fluorescence microscopy movie acquired for a sample in the smectic phase. Only a small fraction of viruses (see Methods) has been labelled with red dyes. The white line indicates the average virus orientation, *i.e.* the nematic director. (B) The resulting differential intensity correlation function (DICF) spectrum (see Fig. S1 for details, ESI†) displays clear anisotropy. For analysis the DICF is averaged over 15° wide sectors, as indicated for the parallel and perpendicular directions.

### 2.2 Differential dynamic microscopy (DDM)

The same set of data as described in the section above are used for DDM analysis. As the conditions are not optimal in this case due to weak signal-to-noise ratio, a further specific sample is prepared in the smectic range with a much higher fraction of red labelled viruses (around $1:10^3$) and for which very long movies (up to 22 min) are recorded.

DDM characterises the spatio-temporal density fluctuations within a sample by computing the differential intensity correlation function (DICF), also known as the image structure function:[11]

$$g(\mathbf{q},\tau) = \langle |\tilde{I}(\mathbf{q},t+\tau) - \tilde{I}(\mathbf{q},t)|^2 \rangle_t \quad (1)$$

with $\tilde{I}(\mathbf{q},t)$ the (2D) Fourier transform of the image intensity $I(\mathbf{r},t)$ and lag time $\tau$. By assuming that the intensity fluctuations in the image are proportional to fluctuations in sample density ($\Delta I \propto \Delta \rho$), the DICF can be written as[12]

$$g(\mathbf{q},\tau) = A(\mathbf{q})[1 - f(\mathbf{q},\tau)] + B(\mathbf{q}), \quad (2)$$

where $A(\mathbf{q})$ characterises the signal amplitude and $B(\mathbf{q})$ accounts for uncorrelated background noise. Here $f(\mathbf{q},\tau)$, often referred to as the intermediate scattering function (ISF), is the $\mathbf{q}$-Fourier component of the probability of the particle displacements, $\delta \mathbf{r} = \mathbf{r}_j(t+\tau) - \mathbf{r}_j(t)$,

$$f(\mathbf{q},\tau) = \langle e^{i\mathbf{q}\cdot\delta\mathbf{r}} \rangle_{j,t}, \quad (3)$$

with brackets denoting averages over all particles $j$ and times $t$. In most of the 'standard' DDM experiments[11,34] the underlying dynamics are isotropic and the ISF (and thus the DICF) only depends on $|\mathbf{q}| = q$. The 2-dimensional DICF can then be reduced to $g(q,\tau)$ by taking azimuthal averages, and by fitting with a suitable model ISF function, the dynamics can be quantified (Fig. S2, ESI†). DDM can also be used to characterise anisotropic motion, as encountered in the nematic and smectic phases of our virus suspensions. Here we follow the analysis scheme introduced in previous publications[12,15] where the DICF is subdivided into angular sectors, which get analysed separately in order to take the directional variation of the dynamics into account.

For anisotropic diffusive motion the ISF is then given by

$$f(\mathbf{q},\tau) = e^{-\Gamma(q,\theta)\tau}, \quad (4)$$

where $\theta$ is the angle between $\mathbf{q}$ and the nematic director (mean particle orientation), as shown in Fig. 1. The decay rate $\Gamma(q,\theta) = D(\theta)q^2$ is linked to the direction dependent diffusion coefficient $D(\theta) = D_\parallel \cos^2\theta + D_\perp \sin^2\theta$. The dynamics of most of our experimental data is well described by this simple ISF, but for the densest sample in the smectic range a stretched exponential model provides a more consistent fit to extract the underlying dynamics in the parallel direction:

$$f(\mathbf{q},\tau) = e^{-(\Gamma(q,\theta)\tau)^{\beta(q,\theta)}}, \quad (5)$$

where $\beta$ is the stretch exponent. For $\beta = 1$ this reduces to the diffusive model, whereas $\beta < 1$ corresponds to sub-diffusive motion.

To apply the DDM analysis approach as defined in eqn (4) and (5), one has to assume that all particles throughout the analysed region of interest are aligned in the same direction, which stays constant over time. In practice, virus suspensions only have finite sized domains of unidirectional alignment (uniform director). Therefore we first visually identify square sub-regions of interest (ROI) of $400 \times 400$ pixels which exhibit good alignment and which are free of topological defects within the originally $640 \times 1080$ pixels movie frames. For each of these sub-regions we calculate the 2D DICF, determine the direction of slowest dynamics and then use 12 angular sectors of $15°$ width centred on this direction for further analysis. Averaging these sets of $g(q,\theta,\tau)$ for several ROIs increases the statistics and therefore improves the signal-to-noise ratio.

In order to reliably extract diffusion coefficients from data which are fairly noisy and also affected by photo-bleaching, we first fit the DICFs based on eqn (2) and (4) independently for each $q$ to extract $\Gamma(q,\theta)$. From this we estimate the long time diffusion coefficient $D(\theta)$ for each angular sector by fitting

$$\Gamma(q,\theta) = \Gamma_0(\theta) + D(\theta)q^2, \quad (6)$$

over a suitable range of low $q$ values. The (small) offset $\Gamma_0$ helps to account for the extra decay component introduced by photo-bleaching of our fluorescent samples as well as the difficulty to quantify decays at very low $q$ ($q \to 0$), when $1/\Gamma$ starts to exceed the movie duration. This offset becomes relevant for the very slow dynamics of the dense nematic and smectic phase samples (see ESI† for details).

The signal amplitude $A(\mathbf{q})$ is often simply treated as a fitting parameter in eqn (2), but it can be used to extract additional information about the sample.[36–38] For monodisperse particles in shape and size we can approximate the amplitude as[12,38]

$$A(\mathbf{q}) \approx N\kappa^2 |OTF(q)|^2 \langle |P(\mathbf{q})|^2 \rangle_{j,k} S(\mathbf{q}), \quad (7)$$

where $N$ is the number of particles in the field of view, $\kappa$ is the overall intensity contrast of a single particle, $OTF(q)$ is the optical transfer function of the objective (assumed to be rotationally symmetric), $P(\mathbf{q})$ is the particle's form factor and $S(\mathbf{q})$ is the sample's static structure factor. For simplicity we have already neglected effects of the objective's finite depth of field.[38,39] Note that the form factor contribution $\langle |P(\mathbf{q})|^2 \rangle_{j,k}$ gets averaged over all particles in the field of view and all time points included in the DDM analysis. Therefore $A(\mathbf{q})$ remains isotropic even for very anisotropic particles provided they are free to sufficiently explore different orientations during the duration of the analysed movies, as in the disordered liquid phase. To date, $A(\mathbf{q})$ has therefore been considered as isotropic, but in the nematic and smectic phases of virus suspensions as discussed here, its anisotropy becomes clearly evident and has to be taken into account.

Note that in our experiments only the small fraction of viruses which are fluorescently labelled ('tracers') contributes to the DDM signal, so that we are only imaging the dilute subset of viruses and thus there is no effect from the sample structure factor ($S(\mathbf{q}) \sim 1$). This also implies that dynamical correlations between different tracers can be ignored and therefore the diffusion coefficient measured by DDM is the self-diffusion coefficient of the tracers.

# 3 Results

Experimental data are acquired for a wide range of concentrations and the resulting movies are analysed using single particle tracking and differential dynamic microscopy as detailed in the previous sections. We found that we could successfully extract the diffusion coefficients across the whole range of packing fractions as shown in Fig. 2. In the following we will present the data for different regimes in more detail.

## 3.1 Dilute isotropic liquid samples

The translational diffusion coefficient $D_0$ of a Brownian rod in the isotropic phase is the geometrical average of the diffusion coefficients along, $D_{0,\parallel}$, and perpendicular, $D_{0,\perp}$ to the rod long axis: $D_0 = \frac{1}{3}(D_{0,\parallel} + 2D_{0,\perp})$. In the limit of infinite dilution, the rods move without interacting with each other and only experience hydrodynamic friction with the solvent. This leads to $D_{0,\parallel/\perp} = k_B T/\gamma_{\parallel/\perp}$ where $\gamma_{\parallel/\perp}$ are the friction coefficients. For slender and stiff rods, $\gamma_{\parallel/\perp}$ are given by $\gamma_\parallel = \frac{2\pi\eta_0 L}{\ln\{L/d\}}$ and $\gamma_\perp = 2\gamma_\parallel$ with $L$ and $d$ the rod length and diameter respectively, and $\eta_0 = 0.89 \times 10^{-3}$ Pa s the water viscosity at room temperature (25 °C).[40] Numerically, by neglecting both the M13KE virus semi-flexibility and the electrostatic interactions, this gives $D_0^{\text{theo}} = 2.4$ μm$^2$ s$^{-1}$ to be compared to the experimental ones, $D_0^{\text{SPT}} = 2.2$ μm$^2$ s$^{-1}$ and $D_0^{\text{DDM}} = 2.3$ μm$^2$ s$^{-1}$, obtained by SPT and DDM, respectively. Overall, an excellent agreement is found between the two experimental values and the theoretical one, validating our experimental setup and our methods for data analysis.

## 3.2 Semi-dilute regime

As the virus concentration is increased from the dilute liquid phase, the dynamics starts to slow down markedly while still remaining isotropic (Fig. 2 and 3). This can be seen in the drop of the diffusion coefficient as shown in Fig. 3, where the logarithmic concentration axis helps to distinguish two regimes. In the dilute regime the diffusion coefficient shows nearly no concentration dependence, remaining close to the one at infinite dilution $D_0$. As the concentration is increased the semi-flexible rods start to interact with each other, leading to a strong slowdown in the semi-dilute regime. The transition between the two regimes appears significantly above the overlap concentration $C^* \propto 1/L_e^3 \simeq 0.07$ mg mL$^{-1}$ (Fig. 3).

Doi & Edwards seminal theory for stiff rods in the needle-like limit predicts $D/D_0$ to be a function of the rod concentration above the overlap concentration $C^*$, and to converge to $D/D_0 = 0.5$ near the isotropic-to-nematic transition.[42] The underlying physical mechanism assumes that rods are caged in tubes set by the surrounding rods, hindering their lateral mobility without major effect on their longitudinal motion. Recent experimental evidence supports the existence of such tubes,[43] which have been found to form quite deeply in the semi-dilute regime in agreement with the results shown in Fig. 3.

Rotational diffusion information is essential for testing more advanced theories,[43,44] but it is challenging to extract experimentally due to limitations in signal-to-noise ratio and frame rate of our movies.[45–47] Hence, a heuristic approach developed by Cush & Russo is employed,[8] for which the exponential fit properly accounts for many sets of experimental data. This approach effectively describes the dynamics within the isotropic liquid phase, for which the diffusion coefficient can be reasonably fit to an exponential decay, reaching $D/D_0$ = 0.13 for the highest concentrations before the isotropic-to-nematic (I–N) transition (Fig. 3).

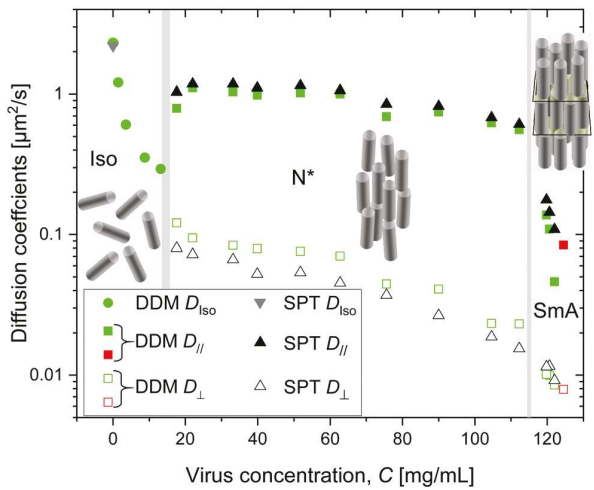

Fig. 2 Isotropic and anisotropic diffusion coefficients determined by differential dynamic microscopy (DDM) and single particle tracking (SPT) measured for the full range of M13KE virus concentrations and organisa-tions: isotropic (Iso), (chiral) nematic (N*) and smectic-A (SmA) phases. The first order phase transitions are indicated by grey areas. For DDM, green and red symbols correspond to samples doped with green and red-labelled fluorescent viruses, respectively.

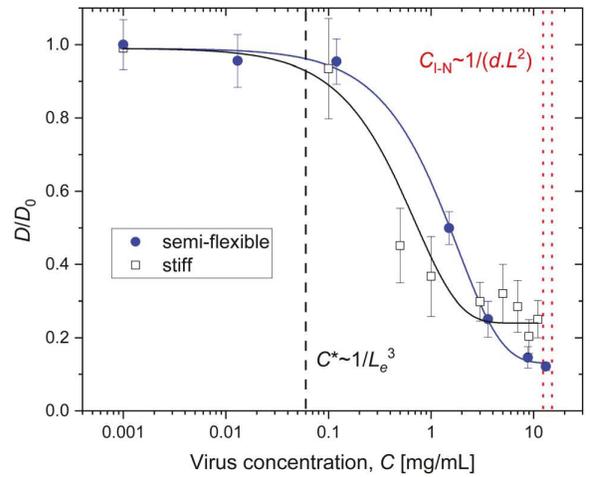

Fig. 3 Diffusion coefficients normalised by the one at infinite dilution ($D_0$) measured in the isotropic liquid phase of virus suspensions by DDM for both semi-flexible M13KE (full blue circle) and stiff Y21M (black open squares) viral rods. The blue and black lines are corresponding single exponential fits over the full range of isotropic phase.[8] The transition between the dilute and the semi-dilute regime in the isotropic liquid phase is shown to be significantly above the overlap concentration $C^*$ indicated by a black dashed line. The isotropic-to-nematic transition given by $C_{I-N} \propto 1/(dL^2)$ according to Onsager theory[41] is shown by a red dotted line for both viruses. Error bars correspond to the normalised standard deviations obtained from fitting the different movies.

This significant decrease of diffusivity below $D/D_0$ = 0.5 has previously been assumed to stem either from the finite rod diameter[44] or from the semi-flexibility[8] encountered in many experimental systems: a finite rod diameter reduces diffusivity by promoting head-on collisions within the tubes, whereas rod semi-flexibility promotes more entanglement in the semi-dilute regime. To gain some insight on the relative importance of these two contributions, we leverage the versatility of filamentous bacteriophages to work with a mutant, Y21M, which is a stiff rod having a higher persistence length $L_p/L \sim 11$ (see Methods). The diffusion coefficients of Y21M suspensions are shown in Fig. 3 for the full range of the isotropic liquid phase. The data for these stiffer viruses show a similar trend to the semi-flexible rods. However there is a reduction in the transition concentration associated with the tube formation as well as a slight increase in the $D/D_0$ near the I–N transition. The value of $D/D_0$ = 0.24 of the stiffer viruses is still significantly lower than the predicted 0.5, highlighting that the finite rod diameter plays a key role in reducing diffusivity.[43,44] Note that the further reduction to $D/D_0$ = 0.13 observed for the semi-flexible viruses may arise from rod flexibility, even though it can be also rationalised in terms of head-on collisions, as flexibility leads to an increased effective diameter due to thermal fluctuations of the virus backbone.[31] The observation that the semi-dilute regime is initiated at lower concentrations for stiff rods is consistent with the entropic role played by rod flexibility for disentanglement from the tubes. Overall, besides the primary impact of finite rod diameter, we show using DDM that rod flexibility significantly influences the decrease of translational diffusion in the semi-dilute regime of the isotropic phase of rod-like virus suspensions.

### 3.3 Nematic phase

By increasing further the concentration, the virus suspensions self-organise into the first liquid crystalline state, i.e. the chiral nematic phase, where rods exhibit long-range orientational order. As the cholesteric periodicity in the range of 20 to 200 μm far exceeds the typical length scales associated with the system,[48,49] our rod suspension can be considered locally as a nematic phase from the point of view of the dynamics. The anisotropy of the mesophase results in an anisotropy of its dynamics with two diffusion coefficients $D_\parallel$ and $D_\perp$ in the direction parallel and perpendicular to the long rod axis, respectively.[10] Single particle tracking shows that, in the nematic phase, the motion remains nearly diffusive (Fig. S6, ESI†), with the parallel diffusion prevailing with respect to the perpendicular one, as $D_\parallel/D_\perp \simeq 10$.[10,23] Such results are obtained by fitting the MSD with a linear function of time with a constant offset – independent of concentration – introduced to account for the finite instrumental resolution (see ESI†). Note that a more sensitive analysis for probing the instantaneous dynamics of such systems is to study the time evolution of the probability density function, or self-Van Hove functions, as shown recently.[24] DDM analysis confirms these findings, with DICF's well fitted using a simple diffusive model (eqn (4)) and the decay rates $\Gamma(q)$ showing the expected $q^2$ scaling in both directions (Fig. S3B, ESI†). Stretched exponential fits (eqn (5)) return stretch exponents close to 1, reconfirming that the motion is still very close to a diffusive regime. DDM and SPT results also agree well quantitatively as seen from the extracted long-time diffusion coefficients shown in Fig. 2. However, a slight systematic deviation, that is lower $D_\parallel$ and higher $D_\perp$, is found for diffusion coefficients obtained by DDM as compared to SPT. This can be rationalised by the fact that finite angular sectors of 15° width (see Methods) are used for DDM analysis. More importantly, DDM is intrinsically a statistically average method which quantifies the dynamics with respect to a fixed direction within the imaged region of interest. Unlike SPT, where the analysis is performed within the frame of individual rods, our DDM analysis does not account for director fluctuations.

### 3.4 Smectic-A phase

By further increasing the virus concentration, the smectic-A phase appears and this lamellar organisation can be considered as a one-dimensional stack of liquid slabs.[50] In this phase, the rod-like viruses behave mostly as Brownian particles in a one-dimensional potential.[22,32,35] This results in a self-diffusion taking place preferentially in the direction normal to the smectic layers, and occurring by quasi-quantised steps of one rod length as evidenced by SPT.[22,23] The long-time diffusion coefficient can be readily extracted from the MSD derived from SPT data. The MSDs also display clear signatures of sub-diffusive motion at short delay times $\tau$ in the parallel direction, whereas the motion in perpendicular direction still appears to be purely diffusive (Fig. S6, ESI†). DDM analysis of the same data can also extract long-time diffusion coefficients, which are in good agreement with the tracking data (Fig. 2 and Fig. S3C, ESI†). However, although there is some indication that the parallel motion is better fitted by the stretched model (eqn (5)), the DDM data is too noisy to reliably detect any deviation from purely diffusive dynamics.

We therefore prepared an additional sample in the smectic-A phase with a higher fraction of labelled viral particles, for which very long movies up to 22 min at 20 fps have been recorded (see Methods). The vastly increased DDM signal levels (Fig. S7, ESI†) combined with improved statistics make it possible to reliably quantify the dynamics across a much larger $q$-range using stretched exponential fitting (Fig. 4 and Fig. S4, ESI†). In the perpendicular direction the decay rate $\Gamma$ scales with $q^2$ and the stretch exponents remains very close to 1 across the whole $q$ range. This confirms that the motion is purely diffusive, in very good agreement with SPT analysis. But in the parallel direction two regimes can now be distinguished: (1) at low $q$ (i.e. $q < 2$ μm$^{-1}$) the dynamics is diffusive with $\beta_\parallel \approx 1$ and $\Gamma_\parallel \propto q^2$ (Fig. 4). This diffusive regime can be interpreted as the long time behaviour of a 1D random walker, for which the unit displacement or step is the hopping event of a virus between two adjacent layers. Probing this regime requires to observe multiple jump events to fully capture the underlying dynamics at long times. It is worth mentioning that this provides the first experimental evidence of this very long time diffusive behaviour in the smectic phase, which is challenging to detect with SPT as most particles become untrackable over extended periods of time. We estimate the long-time diffusion coefficient as shown in Fig. 2 from $\Gamma$ for $q \leq 2 \ \mu m^{-1}$ (where $\beta \geq 0.93$). (2) At larger wavevectors ($q \gtrsim 3 \ \mu m^{-1}$) the significant drop in stretch exponent ($\beta_{\parallel} \lesssim 0.8$; Fig. 4A) indicates more complex dynamics such as the presence of multiple characteristic time scales.[15] This is reinforced by the non-monotonic behaviour in decay rate $\Gamma_{\parallel}$, indicating that there are at least two different processes with different $q$-scaling. Closer inspection of the DICF curves indeed reveals the emergence of a separate slower decay component as $q$ approaches the value associated with the smectic layer thickness $2\pi/L_{layer} \approx 6.3 \ \mu m^{-1}$. Scaling the delay time by $q^2$ as shown in Fig. 4C helps to highlight these two separate time scales and suggests that the faster initial decay is diffusive. The slower second decay stems from the transient trapping of the particles within the smectic layers, with the decay rate $\Gamma$ decreasing as the associated length scales $\ell = 2\pi/q$ approach the layer thickness $L_{layer}$. Although it is possible to extract two distinct decay rates by fitting the DICF for $q \gtrsim 4 \ \mu m^{-1}$ with a more complex fit (Fig. S4 and S5, ESI†), quantitative interpretation would require a theoretical expression of the intermediate scattering function and faster frame rates to allow better characterisation at short delay time. This is beyond the scope of the current manuscript. Also note that the minimum in the form factor (see Section 3.5) makes it difficult for DDM to accurately probe the dynamics very close to the layer length scale, as $L_{layer} \simeq L$.

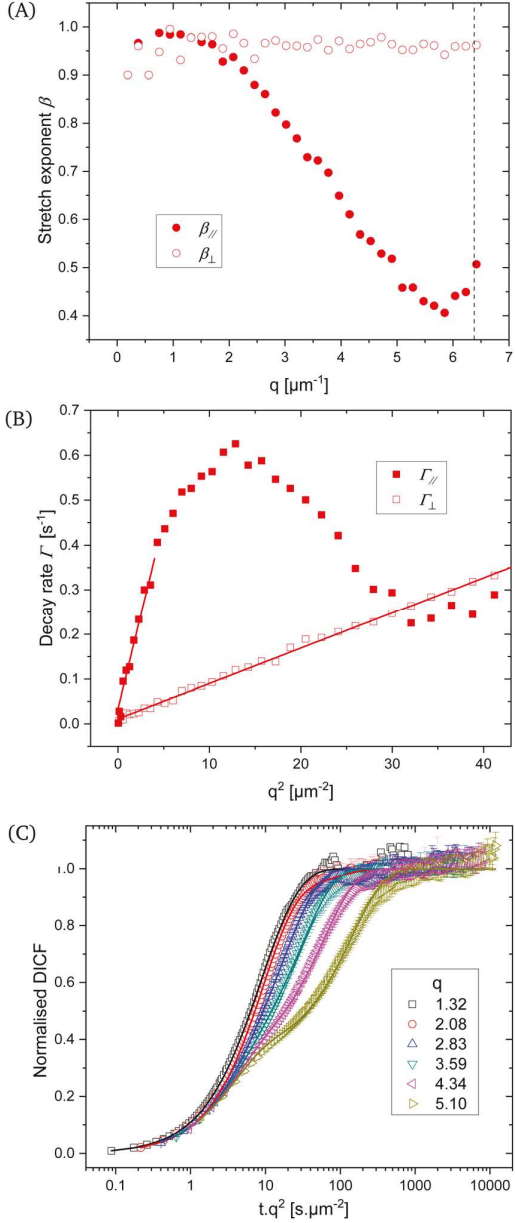

Fig. 4 (A) Stretch exponential fit parameters $\beta$ using eqn (5) for smectic-A sample with red fluorescent dyes optimised for DDM analysis. Full and open symbols correspond to the directions parallel and perpendicular to the director, respectively. For parallel motion, a diffusive behaviour is reached at long time, corresponding to $q < 2 \ \mu m^{-1}$. The vertical dashed line indicates the first minimum of the form factor (see main text and Fig. 5). (B) The anisotropic decay rates $\Gamma(q^2)$ are used to determine the long-time diffusion coefficients (see eqn (6)) by linear fits within the $q$-ranges where $\beta \geq 0.93$, as indicated by solid lines. Resulting $D_{\parallel,\perp}$ are shown in Fig. 2 by the two red symbols. (C) Normalised DICF (open symbols) for parallel motion reveal two distinct processes for $q > 4 \ \mu m^{-1}$ as shown by double-decay fits (solid lines) for the two highest $q$ shown here (see Fig. S5 for fit values and Fig. S4 for perpendicular motion, ESI†). For $q < 4 \ \mu m^{-1}$, a stretched exponential fit properly accounts for the single dynamic process in the smectic phase.

### 3.5 Form factor

The bodies of our tracer viruses are uniformly labelled with dyes, making their rod-like shape clearly visible in our high magnification movies (Fig. 1). The form factor of our viruses should therefore be well described by that of a cylindrical particle of length $L$ and radius $r$, given by[51]

$$P(q, \phi) = 2\frac{J_1(qr\sin\phi)}{qr\sin\phi} \times \frac{\sin(q(L/2)\cos\phi)}{q(L/2)\cos\phi}, \quad (8)$$

where $\phi$ is the angle between **q** and the cylinder axis and $J_1$ the Bessel function of the first kind. For the direction parallel to the main rod axis, $\phi = 0$ and eqn (8) reduces to $P(q, \phi = 0) \propto \frac{\sin(q(L/2))}{q(L/2)}$ whereas in the perpendicular direction $\phi = 90°$, this yields to $P(q, \phi = 90°) \propto 2\frac{J_1(qr)}{qr}$, which for a slender rod stays $\approx 1$ across the accessible $q$ range. This means that DDM amplitude signal in the perpendicular direction (as shown in Fig. S7, ESI†) mostly accounts for the optical transfer function of our microscope (see eqn (7)), and can then be used to extract the square of the form factor of our particle in the parallel direction, as described in detail in ESI† and shown in Fig. 5.

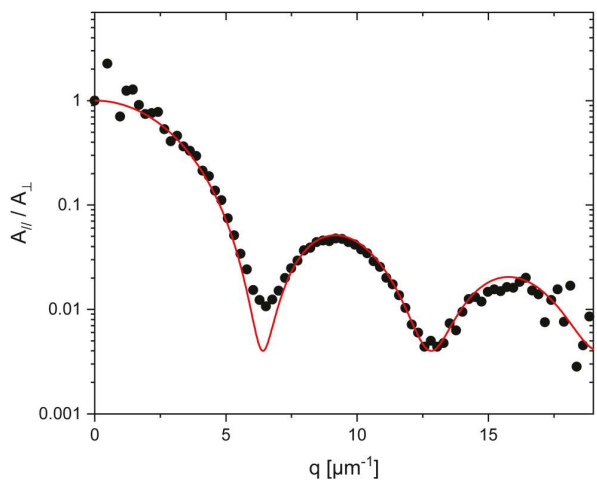

Fig. 5 Experimental DDM signal amplitude ratio $A_\parallel/A_\perp$ (blacks dots) is accounted by the square of the form factor of a cylinder (red line). The numerical fit gives a cylinder length $L = 0.98$ μm. The value is in very good agreement with the virus contour length of 1 μm.

This agrees very well with the expected $q$-dependence, and the position of the minima can be used to provide an independent estimate of the virus length $L$ (Fig. 5). Note that this strong asymmetry of the virus form factor makes the interpretation of DDM data for higher $q$-values more difficult. For small wavenumbers ($q \lesssim 2$ μm$^{-1}$) the form factor stays close to unity and therefore the contributions to the DDM signal are essentially independent of the direction of motion relative to the particle axis. But for higher $q$ values the contribution from motion parallel to the virus axis drops quickly as it approaches $q_L = 2\pi/L$. In the present case, small perpendicular components are still detected due to the finite sector size and ultimately dominates the dynamic signal near the form factor minima.

This constitutes an intrinsic difficulty to reveal the dynamics of rod-shaped particles in the direction of their long axis for any reciprocal space method, such as DDM and DLS for instance, involving the form factor of a cylinder. In case of DDM this could be overcome by only labelling the virus centre of mass with dyes, thereby removing the anisotropy of the form factor.

## 4 Conclusions

In this study, we have investigated the dynamics of self-assembled suspensions of rod-like fd virus colloids across a wide range of concentrations using single particle tracking (SPT) and differential dynamic microscopy (DDM). Both techniques have provided valuable insights into the anisotropic diffusion of the viral rods in various self-assembled states, including dilute and dense isotropic liquids, nematic and smectic phases.

SPT, a well-established technique, has proved to be a powerful and accurate tool to directly measure the anisotropy of diffusion coefficients, particularly in the nematic and smectic phases. SPT is able to capture detailed information about the dynamic behaviour in the direct space, as the hopping-type diffusion in the smectic phase. However, obtaining high statistics with SPT can be challenging. In contrast, DDM, an ensemble-averaged technique, is relatively fast and efficient to deliver a rapid characterisation of the sample-averaged dynamics. While DDM results have shown good quantitative agreement with diffusion coefficients extracted from SPT across all concentrations, it struggles to account for local variations within the analysed area. This limitation leads to small systematic deviations for nematic and smectic samples compared to SPT, whose analysis can be advantageously performed in the frame of the particle. A major strength of DDM lies in its ability to achieve good statistics by simultaneously imaging many tracer particles over extended time periods. In anisotropic liquid crystalline phases, this advantage can be exploited by increasing the tracer concentration, as successfully demonstrated here for a smectic sample. By analysing the $q$-dependence of the DDM signal, we have been able to reveal a diffusive regime at long times in the smectic phase for motion parallel to the normal of the smectic layers. This finding is challenging to access with SPT due to limitations in tracking particles over extended periods. For isotropic samples in the semi-dilute regime, we have explicitly demonstrated that the decrease of the diffusivity in the semi-dilute regime depends on the particle flexibility, beyond the primary influence of finite particle diameter. DDM has been shown to be an efficient method in this semi-dilute regime, which, for more detailed studies, could be further exploited by using much larger fields of view, lower magnification imaging and more advanced data processing[47] to potentially provide insights into rotational diffusion.

Overall, our study emphasises the importance of considering both the underlying physics of the system and the limitations of each technique to gain a comprehensive understanding of the complex dynamics within anisotropic colloidal systems.

## Author contributions

Eric Grelet: conceptualisation, formal analysis, funding acquisition, resources, investigation, methodology, software, visualisation, writing – original draft, writing – review and editing. Vincent A. Martinez: methodology, writing – review and editing. Jochen Arlt: conceptualisation, formal analysis, funding acquisition, methodology, software, visualisation, writing – original draft, writing – review and editing.

## Data availability

Data relevant to this publication are available upon request from the corresponding authors as well as on the following DataShare: **https://doi.org/10.7488/ds/7850**.

## Conflicts of interest

There are no conflicts to declare.

## Acknowledgements

We acknowledge funding from the European Unions Horizon 2020 research and innovation programme under grant agreement no. 731019 (EUSMI). We thank A. Repula for his contribution to SPT analysis, and L. Alvarez for the raw data in the isotropic phase of the stiff mutant.

## Notes and references

# ELECTRONIC SUPPLEMENTARY INFORMATION

## for:

## Self-diffusion in Isotropic and Liquid Crystalline Phases of fd Virus Colloidal Rods: a Combined Single Particle Tracking and Differential Dynamic Microscopy Study

**Details for extracting long time diffusion coefficients with DDM**

As briefly outlined in the main text, anisotropic DDM analysis was performed on well aligned sub-regions selected through visual inspection. The direction of slowest dynamics is easily identified as a 'low signal line' in the 2D DICF at short delay times (Fig. S1A). This corresponds to the direction perpendicular to the rod axis, and was experimentally found to be consistent with the visually identified rod direction in the fluorescence movies (*i.e.* perpendicular to it). We then used 12 angular sectors of $\Delta\theta = 15°$ width centred on this direction for further analysis. Averaging these sets of $g(q, \theta, \tau)$ for several ROIs helped to increase the signal-to-noise ratio.

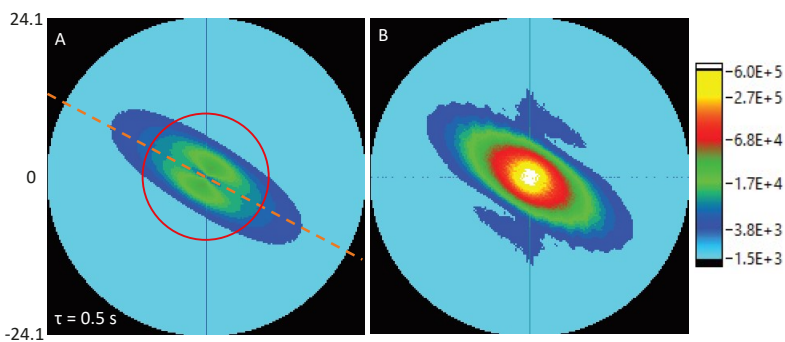

Fig. S1 DICF spectra for a densely labelled sample in the smectic phase. (A) After a short delay time the direction of slow dynamics can be easily identified (dashed orange line). (B) After a very long delay, where the signal amplitude can be used to estimate the square of the virus' form factor. The $q$ range of both spectra is $\pm 24.1\,\mu m^{-1}$ and the DDM signal amplitude is visualised using a logarithmic colour scale as indicated. Note that for the green labelled samples we could only analyse a much smaller $q$ range as indicated by the red circle.

Representative DICF's are shown in Fig. S2 for both isotropic and nematic samples.

In the low $q$-limit, the long time diffusion coefficient $D(\theta)$ was estimated by fitting

$$\Gamma(q,\theta) = \Gamma_0(\theta) + D(\theta)q^2, \tag{S1}$$

over a suitable range of $q$ values (Fig. S3).

For isotropic samples there was no need for sector based analysis, and the decay constant initially scales linearly with $q^2$, as expected for purely diffusive dynamics. But as can be seen from Fig. S3A, with increasing density there is clear deviation from this initially linear trend for large $q^2$. Here we used a linear fit up to $q^2 = 4\,\mu m^{-2}$ to estimate the long time diffusion coefficient.

For nematic samples the decay rates vary with $\theta$, *i.e.* the direction of motion within the observation plane. Fig. S3B shows some representative data for motion in the parallel and perpendicular direction for three different concentrations. The slow dynamics in the perpendicular direction shows diffusive scaling over the complete $q^2$ range, whereas the decay rates in the parallel direction can only be extracted up to about $10\,\mu m^{-2}$. This shorter range is due to a drop in signal levels caused by the more pronounced effect of the form factor while the decay rate approaches the frame rate of the movie (33 fps). In the smectic phase (Fig. S3C) the overall dynamics is even slower and the scaling in the parallel direction fails at much lower $q^2$ values. The small offset $\Gamma_0$ becomes clearly noticeable here.

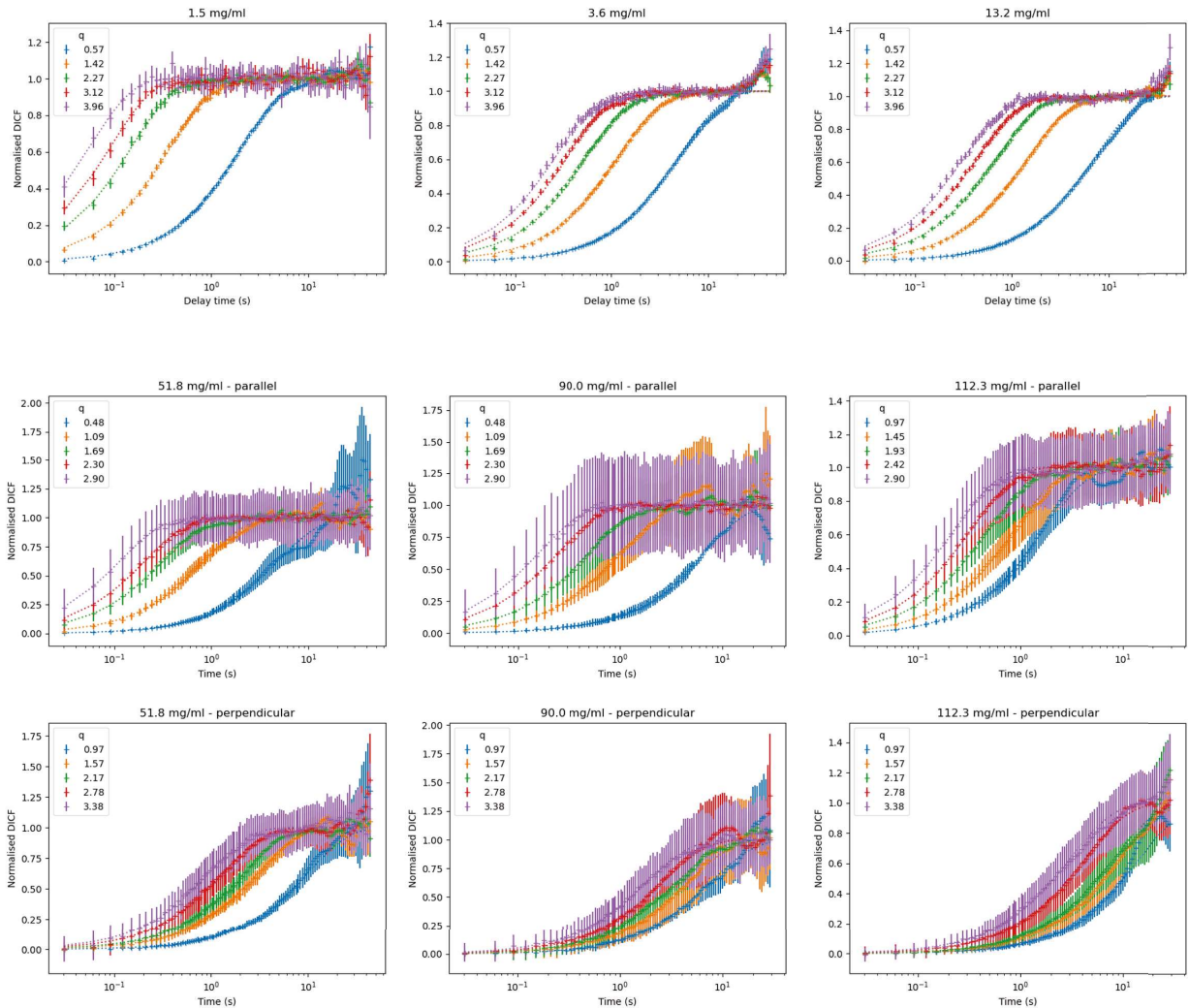

Fig. S2 Normalised DICFs for different sample concentrations for a range of $q$-values (symbols) together with fit curves assuming a diffusive ISF (dotted lines). The top row shows DICFs for three isotropic samples: as the data can be averaged over all azimuthal angles the experimental data has relatively small error bars. For nematic samples the DICF becomes direction dependent: The second row shows the (faster) dynamics parallel to the director and the third row the corresponding data in the perpendicular direction. Note that the data are noisier due to the poorer statistics afforded by averaging over $15°$ angular sectors.

For both the nematic and smectic phases we extracted the long time diffusion coefficient in the parallel direction by fitting Eq. S1 up to $q^2 = 4\,\mu\text{m}^{-2}$ and used the whole range for the perpendicular direction.

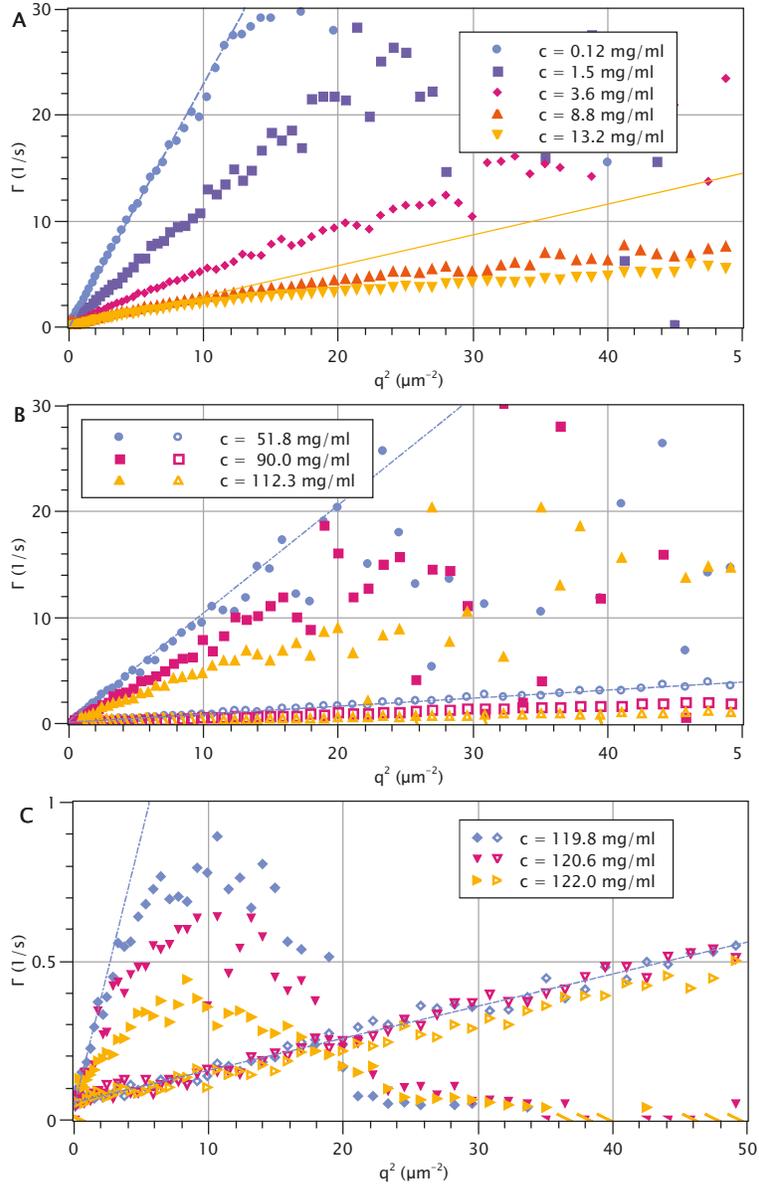

Fig. S3 Decay constants $\Gamma$ plotted against $q^2$ for several sample concentrations within the (A) isotropic, (B) nematic and (C) smectic phase. In panels (B) & (C) the filled symbols are for motion parallel to the virus long axis and the open symbols for the much slower motion in the perpendicular direction.

**DICFs for red labelled smectic sample**

The improved contrast and increased density of labelled viruses allows us to extract much cleaner DICFs, as shown in Fig. S4. In the perpendicular direction smectic samples retain simple exponential ISFs for all $q$ and the DICFs collapse when plotted against delay time scaled by $q^2$, confirming that its dynamics is purely diffusive. But in the parallel direction we can now clearly see deviations a simple exponential ISF. A stretched exponential fit captures the dominant dynamics well (Fig. S4 B&E), however for $q \gtrsim 4\,\mu\text{m}^{-1}$, close to length scale $1/q$ associated with the layer spacing, two distinct decay processes can be visually identified. Using a double-exponential fit,

$$f(q,\tau) = (1-\alpha)\,e^{-D_{\text{fast}}q^2\tau} + \alpha\,e^{-D_{\text{fast}}q^2\tau}, \tag{S2}$$

both processes are captured well for $4\,\mu\text{m}^{-1} < q < 5.1\,\mu\text{m}^{-1}$, where the faster process has $D_{\text{fast}} \approx 0.26(5)\,\mu\text{m}^2/\text{s}$ and the dominant slower process $D_{\text{slow}} \approx 0.014(6)\,\mu\text{m}^2/\text{s}$, with a fractional contribution $\alpha \approx 0.65$ to the overall signal (Fig. S5). Note that there is no *a priori* reason to assume that these processes are diffusive. However, to extract physical insights from the use of a double-stretched fit model (*i.e.* a superposition of two stretched exponentials) a more systematic study would be needed, as well as access to shorter delay times.

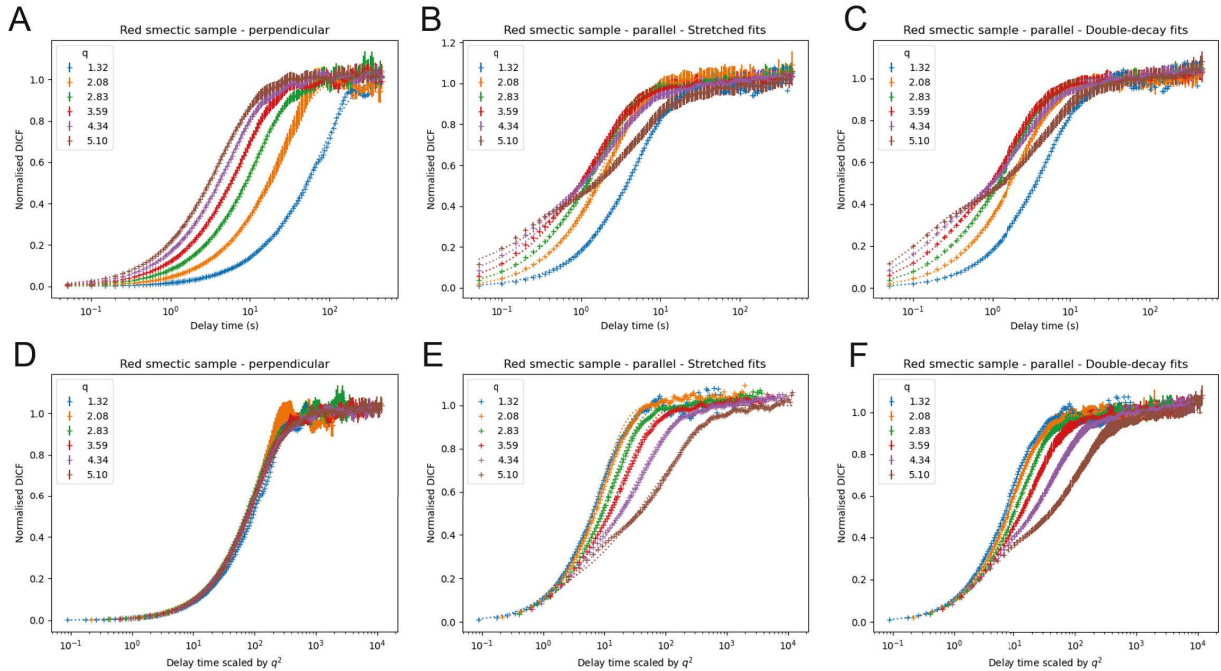

Fig. S4 Normalised DICFs for selected $q$ values for the densely red labelled sample in the smectic phase. Corresponding fit curves are displayed as dotted lines. Top left plot (A) shows DICFs for motion perpendicular to the director. The middle (B) and right (C) plots are for motion parallel to the director together with a stretched fit or double-exponential fit, respectively. The bottom row (D-F) shows the same data but with the delay time scaled by $q^2$. Note that panel (E) does not show error bars to make systematic deviations of stretched fit more visible.

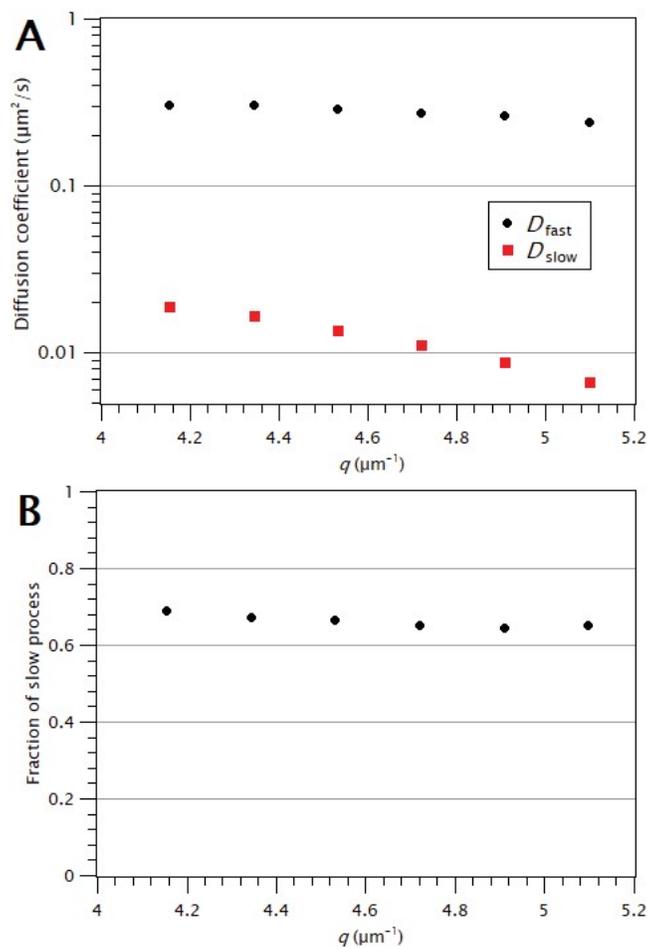

Fig. S5 (A) Diffusion coefficients and (B) fractional contribution of the slow process ($\alpha$, see Eq. S2) for a double-exponential fit to the DICF in the parallel direction.

**Details for extracting long time diffusion coefficients from MSD by SPT**

The mean squared displacements (MSD) averaged over hundreds of traces obtained by SPT (see main text), are fitted to determine the corresponding anisotropic diffusion coefficients $D_\parallel$ and $D_\perp$ in the direction parallel $\parallel$ and perpendicular $\perp$ to the mean rod orientation, respectively.[23] As numerical fit, a power law is used with a constant offset, assumed to be isotropic, and accounting for the finite optical (point spread function) and temporal resolution of the optical setup. As displayed in Fig. S6, most the dynamics of the system is diffusive for the probed time window in the nematic phase for both directions, and in the smectic phase for the perpendicular direction. However, for the smectic phase in the parallel direction, a sub-diffusive regime can be evidenced at short time associated with the hopping-type motion, whereas a diffusive regime is recovered at long time (Red full circle symbols in Fig. S6).

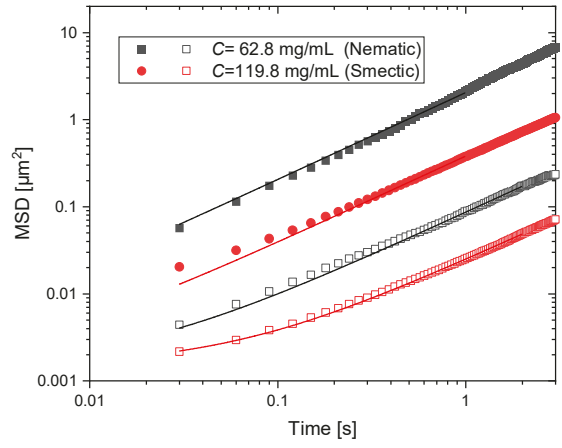

Fig. S6 Mean squared displacements as a function of time for two virus concentrations of $C = 62.8$ mg/mL and $C = 119.8$ mg/mL corresponding to the nematic (black symbols) and smectic (red symbols) liquid crystalline phases, respectively. The filled symbols are for parallel displacements along the virus director and the open symbols are for the perpendicular displacements. The continuous lines are diffusive fits with a fixed offset for both concentrations and directions: $MSD_{\parallel,\perp} = 2D_{\parallel,\perp} t + 0.0015\,\mu m^2$. The small offset accounts for the dynamic localisation uncertainty.

**Estimating the form factor from the DDM signal amplitude**

At large delay times $\tau$ the amplitude of the DICF spectrum as shown in Fig. S1 B reveals the **q**-variation of the signal amplitude $A(\mathbf{q})$. Fig. S7 shows the amplitudes for 12 sectors, with the magenta data points corresponding to motion along the virus axis, $A_\parallel(q)$ at $\theta = 0°$ and the black symbols to motion in the perpendicular direction, $A_\perp(q)$ ($\theta = 90°$). As the physical diameter of the virus is well below the resolution limit of the microscope, i.e. $P_\perp(q) = 1$ throughout the accessible $q$-range, the observed $q$ dependence of $A_\perp$ is only due to the optical transfer function of the objective. By calculating the ratio

$$\frac{A_\parallel(q)}{A_\perp(q)} = \frac{N\kappa^2 |\mathrm{OTF}(q)|^2 |P_\parallel(q)|^2}{N\kappa^2 |\mathrm{OTF}(q)|^2} = |P_\parallel(q)|^2$$

we can therefore get an estimate of the square of the form factor of the virus, as shown in Fig. 5 in the main manuscript.

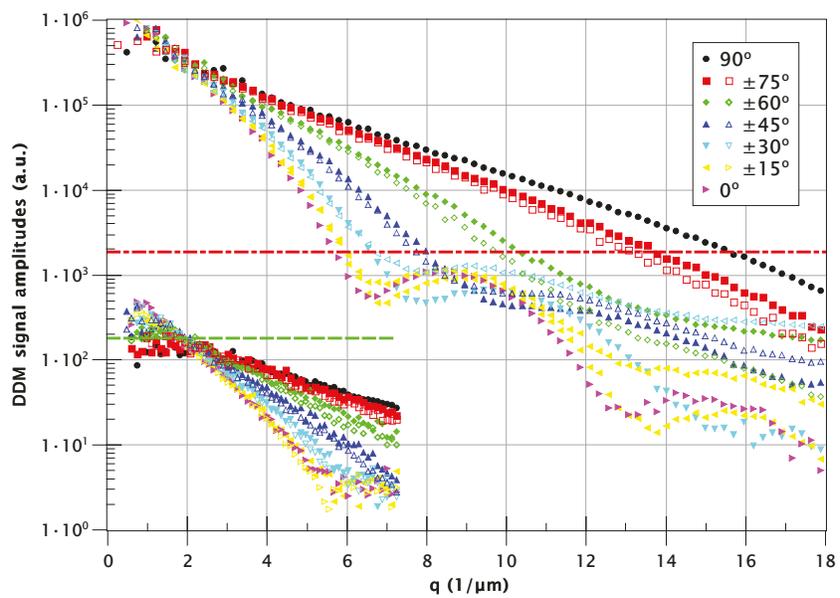

Fig. S7 DDM signal amplitudes as a function of direction relative to the virus director. The signal for motion perpendicular to the virus axis (black dots) drops much more slowly than in the parallel direction (magenta right triangles), for which pronounced minima can be distinguished due to the viruses form factor. The main dataset (with larger signal amplitude spanning a much wider $q$-range) is from a very long movie recorded with red labelled viruses in the smectic phase (see main text). A representative dataset for movies with viruses labelled with green fluorescent dyes, as used for the majority of experimental data presented here, is also shown for comparison. The horizontal dashed lines indicate the average value of background $B(q)$ (see Eq. 2) for the respective datasets.